\documentclass{PoS}
\usepackage{float}
\title{$L_{10}^r$ From a Combined NNLO Lattice, Continuum 
Analysis of the Light Quark $V-A$ Correlator}

\ShortTitle{Combined NNLO Lattice Continuum Analysis of the Light Quark $V-A$
Correlator}

\author{P.A. Boyle,$^a$, L. Del Debbio,$^a$ N. Garron,$^b$,
R.J. Hudspith,$^a$, E. Kerrane$^c$, \speaker{K. Maltman},$^{d,e}$ 
and J.M. Zanotti$^e$
\\
\llap{$^a$}Physics and Astronomy, University of Edinburgh, Edinburgh
EH9 3JZ, UK\\
\llap{$^b$}School of Mathematics, Trinity College, Dublin 2, Ireland\\
\llap{$^c$}Instituto de F\`isica T\`eorica UAM/CSIC, Universidad
Aut\`onoma de Madrid, Cantoblanco E-28049 Madrid, Spain\\
\llap{$^d$}Mathematics and Statistics, York University, Toronto M3J 1P3
Canada\\
\llap{$^e$}CSSM, University of Adelaide, Adelaide 5005 Australia\\
E-mail: \email{paboyle@ph.ed.ac.uk},
\email{ldeldebb@ph.ed.ac.uk}, \email{ngarron@maths.tcd.ie},
\email{s0968574@sms.ed.ac.uk}, \email{eoin.kerrane@gmail.com},
\email{kmaltman@yorku.ca}, \email{james.zanotti@adelaide.edu.au}}

\abstract{A combination of lattice and continuum data for the 
light-quark V-A correlator, supplemented by results from a chiral 
sum-rule analysis of the flavor-breaking flavor $ud$-$us$ V-A correlator 
difference, is shown to make possible a high-precision NNLO determination 
of the renormalized NLO chiral low-energy constant $L_{10}^r$. Key to this 
determination is the ability to simultaneously fix the two combinations of 
NNLO low-energy constants also entering the analysis. With current versions 
of the strange hadronic $\tau$ branching fractions required as input to the
flavor-breaking V-A sum rule, we find $L_{10}^r(m_\rho )\, =\, -0.00346(29)$.
This represents both the best current precision for $L_{10}^r$,
and the first NNLO determination having all errors under full control.}

\FullConference{31st International Symposium on Lattice Field Theory LATTICE
  2013\\
                 July 29 . August 3, 2013\\
                 Mainz, Germany}

\begin{document}

\section{Introduction}
The low-energy effective (chiral) Lagrangian framework encodes, in the 
most general way, the constraints on the light degrees of freedom of the 
symmetries and approximate chiral symmetry of QCD. A key goal in maximizing 
the predictive power of this approach is the completion and/or improvement 
of the determinations of all coefficients (low-energy constants, or LECs) 
accompanying operators allowed by the symmetry arguments out to a given order 
in the chiral expansion. In this paper we focus on an improved, 
next-to-next-to-leading order (NNLO) determination of the renormalized 
next-to-leading order (NLO) LEC $L_{10}^r$. The analysis also yields values 
for two NNLO LEC combinations, the determination of which is crucial to 
reducing the uncertainty on $L_{10}^r$.

Existing determinations of $L_{10}^r$ are based on analyses of the low-$Q^2$ 
behavior of the difference, 
$\Delta\Pi_{V-A}(Q^2)\equiv \Pi_{ud;V-A}^{(0+1)}(Q^2)$, of the
light-quark (flavor $ud$) vector (V) and axial vector (A) correlators.
The spin $J=0,1$ scalar correlators $\Pi_{ud;V/A}^{(J)}(Q^2)$ entering
this definition are defined by the standard decompositions of the 
$ud$ V and A current-current two-point functions. An NLO version of this 
analysis was performed  in Ref.~\cite{dghs98}, with differential hadronic 
$\tau$ decay distribution results for the spectral function, 
$\Delta\rho_{V-A}(s)$, of $\Delta\Pi_{V-A}(Q^2)$ used to fix 
$\Delta\Pi_{V-A}(0)$, and hence $L_{10}^r$, the only free parameter 
in the NLO representation of $\Delta\Pi_{V-A}(0)$. NLO 
analyses were also performed using low-(Euclidean)-$Q^2$ lattice data 
for $\Delta\Pi_{V-A}(Q^2)$~\cite{jlqcdvma,rbcukqcdvma09}. 

The low-energy representations of $\Pi_{ud;V/A}^{(J)}(Q^2)$ are now known 
to NNLO~\cite{abt00}, and yield a value for the coefficient of $L_{10}^r$ 
in the resulting $\Delta\Pi_{V-A}(0)$ representation $\sim 50\%$ larger at 
NNLO than at NLO, raising doubts about the reliability of the earlier NLO 
analyses. The observation (also made for the $ud$ V correlator~\cite{ab06}) 
that the NLO representation fails completely to reproduce the variation 
of $\Delta\Pi_{V-A}(Q^2)$ with $Q^2$~\cite{bgjmp13} confirms these worries. 
Attempts to extend the continuum NLO analysis to NNLO, however, run into 
the problem that the NNLO representation of $\Delta\Pi_{V-A}(0)$ involves 
two additional NNLO LEC combinations, one of which is completely unknown. 
In Ref.~\cite{gapp08}, this combination was assigned a central value zero,
and a rough guess at the error (based only on large-$N_c$ counting, and 
since argued to be rather non-conservative~\cite{bgjmp13}) attributed 
to this choice. This error turns out to completely dominate the uncertainty 
on the resulting determination of $L_{10}^r$~\cite{gapp08} .

In this paper, we address this situation, showing how to combine the 
dispersive determination of $\Delta\Pi_{V-A}(0)$ with lattice data and an 
additional flavor-breaking (FB) chiral sum rule to fix simultaneously 
$L_{10}^r$ and the two NNLO LEC combinations noted above. Section~\ref{sec2} 
provides more detail on the dispersive determination of $\Delta\Pi_{V-A}(0)$, 
the NNLO representation of $\Delta\Pi_{V-A}(Q^2)$, and the problems 
encountered in an NNLO continuum analysis. The use of lattice data and a 
new chiral sum rule, involving $L_{10}^r$ and one of the two NNLO LEC 
combinations, to deal with these problems is then also discussed.
Section~\ref{sec3} presents the results of our analysis.

\section{\label{sec2}Ingredients for the NNLO Determination of $L_{10}^r$}
$\Delta\Pi_{V-A}(Q^2)$ is free of kinematic singularities and satisfies an 
unsubtracted dispersion relation. With $\Delta\overline{\Pi}_{V-A}(Q^2)$ 
and $\Delta\overline{\rho}_{V-A}(s)$ the continuum ($\pi$-pole-subtracted) 
versions of $\Delta\Pi_{V-A}(Q^2)$ and $\Delta\rho_{V-A}(Q^2)$, the 
dispersion relation, written here for spacelike $Q^2\, =\, -q^2\, =\, -s$, 
becomes
\begin{equation}
\Delta\overline{\Pi}_{V-A}(Q^2)\, =\, 
\int_{4m_\pi^2}^\infty\, ds\, {\frac{\Delta\overline{\rho}_{V-A}(s)}
{s+Q^2}}\ .
\label{vmadispreln}\end{equation}
This representation was recently used to generate high-precision
results for $\Delta\overline{\Pi}_{V-A}(Q^2)$~\cite{bgjmp13}. 
The spectral functions, $\Delta\overline{\rho}_{V/A}(s)$, needed on 
the RHS, are accessible up to $s=m_\tau^2$ using OPAL hadronic $\tau$ 
decay data~\cite{opal99}. These have been updated for current branching 
fractions~\cite{dv7}. Above $s=m_\tau^2$, the representation of 
$\Delta\overline{\rho}_{V/A}(s)$ as a sum of 5-loop $D=0$ OPE and 
duality violating contributions, studied in great detail in 
Refs.~\cite{dv7}, was employed. For low $Q^2$, the part of the integral 
involving the experimental spectral data strongly dominates the results for
$\Delta\overline{\Pi}_{V-A}(Q^2)$~\cite{bgjmp13}. 
\unitlength1cm
\begin{figure}[H]
\caption{\label{fig1}Continuum and RBC/UKQCD lattice results for 
$\Delta\bar{\Pi}_{V-A}(Q^2)$ in the low-$Q^2$ region}
\begin{center}
\rotatebox{270}{\mbox{
%\begin{minipage}{10.1cm}
\includegraphics[width=7.4cm,height=9.6cm]{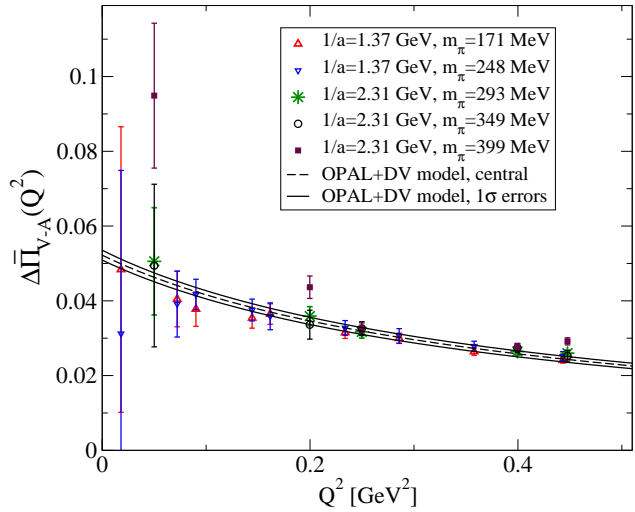}
%\end{minipage}
}}
\end{center}
\end{figure}
For $Q^2>0$, $\Delta\Pi_{V-A}(Q^2)$ can also be determined on the lattice
for a range of $m_u=m_d$ and $m_s$. Ensemble $m_\pi$ and $f_\pi$ values
then yield the corresponding $\Delta\overline{\Pi}_{V-A}(Q^2)$. We work here 
with results for five RBC/UKQCD $n_f=2+1$, domain wall fermion ensembles: 
three, with $m_\pi\, =\, 293$, 
$349$ and $399$ MeV, having inverse lattice spacing $1/a\, =\, 2.31\ GeV$, 
and two, with $m_\pi\, =\, 171$, and $248$ MeV, having $1/a\, =\, 1.37\ GeV$.
Full simulation details may be found in Refs.~\cite{ainv228,ainv137}.

Lattice and continuum results for $\Delta\overline{\Pi}_{V-A}(Q^2)$
are shown, for $Q^2<0.5\  GeV^2$, in Fig.~\ref{fig1}. The continuum
and near-physical-mass, $m_\pi =171\ MeV$, lattice results are in very 
good agreement. This agreement persists to much higher $Q^2$ than shown 
here, suggesting lattice artefacts are safely under control. Lattice 
errors are larger than continuum ones below $Q^2\sim 0.3 \ GeV^2$,
particularly so for $Q^2$ near $0$, where, for Euclidean $Q$, the 
kinematic factors multiplying the scalar correlators in the spin 
decomposition of the two-point functions (hence also the two-point 
functions themselves) vanish in the limit $Q^2\rightarrow 0$.

The NNLO representation of $\Delta\overline{\Pi}_{V-A}(Q^2)$ has the 
form~\cite{abt00}
\begin{equation}
\Delta\overline{\Pi}_{V-A}(Q^2)\, =\, 
c_{10}\, L_{10}^r \, +\, {\cal C}_0^r\, 
+\, {\cal C}_1^r \, +\, c_9\, L_9^r\, +\, R(Q^2)\, -\, 16 C_{87}^r\, Q^2\ ,
\label{nnlodeltapibar}
\end{equation}
where all quantities other than $Q^2$ on the RHS depend on the chiral 
renormalization scale, $\mu$, and
\begin{eqnarray}
&&c_9\, =\, 16\left( 2\mu_\pi +\mu_K\right),\qquad\qquad\qquad
c_{10}\, =\, -8\left( 1-8\mu_\pi -4\mu_K\right)\label{nloleccoefs}\\
&&{\cal C}_0^r\, \equiv\, 32 m_\pi^2 \left( C^r_{12}-C^r_{61}+C^r_{80}\right),
\qquad {\cal C}_1^r\, \equiv\, 32 \left( m_\pi^2+2m_K^2\right)
\left( C^r_{13}-C^r_{62}+C^r_{81}\right)\ ,
\label{nnloleccombodefns}\end{eqnarray}
with $\mu_P\, =\, {\frac{m_P^2}{32\pi^2 f_\pi^2}}\, log\left({\frac
{m_P^2}{\mu^2}}\right)$, and $C_k^r$ the dimensionful renormalized 
NNLO LECs of Ref.~\cite{bce99}. 
${\cal C}_0^r$ and ${\cal C}_1^r$ are leading order in $N_c$ and $1/N_c$ 
suppressed, respectively. The rather lengthy expression for $R(Q^2)$ 
is omitted but easily reconstructed from the results of Sections 4, 6 and 
Appendix B of Ref.~\cite{abt00}. For given $Q^2$, $R(Q^2)$ depends only 
on $\mu$, $f_\pi$ and the pseudoscalar (PS) meson masses~\cite{abt00}. In 
what follows, we define $\hat{R}(Q^2)\, \equiv\,  c_9\, L_9^r \, +\, R(Q^2)$, 
employ $L_9^r(0.77 \ GeV)=0.00593(43)$~\cite{bt02}, and
treat $\hat{R}(Q^2)$ as known once $Q^2$, $\mu$, $f_\pi$ and the PS masses 
are fixed. 

Eq.~(\ref{nnlodeltapibar}) makes evident the problem encountered
in attempting to extend the NLO $L_{10}^r$ determination to NNLO. 
At NLO, with $[R(0)]_{NLO}$ exactly known, the result
$\Delta\overline{\Pi}_{V-A}(0)\, =\, 0.0516(7)$~\cite{bgjmp13} 
translates into a nominally very precise value for $L_{10}^r$. At NNLO, 
however, the constraint becomes
\begin{eqnarray}
&&\Delta\overline{\Pi}_{V-A}(0)\, =\, 0.0516(7) \, =\,
c_{10}\, L_{10}^r \, +\, {\cal C}_0^r\, +\, {\cal C}_1^r\, +\, \hat{R}(0)\ ,
\label{nnlocontpivma0}\end{eqnarray}
and input for ${\cal C}_{0,1}^r$ is required to turn this into a determination
of $L_{10}^r$. Ref.~\cite{gapp08} dealt with this problem using existing 
determinations of $C_{12}^r$~\cite{jop04} and $C_{61}^r$~\cite{dk00,km06}, 
a resonance ChPT (RChPT) estimate for $C_{80}^r$~\cite{up08}, and a 
large-$N_c$-motivated guess for ${\cal C}_1^r$.
The $\sim 26$ enhancement of the mass-dependent factor
in ${\cal C}_1^r$ relative to that in ${\cal C}_0^r$ more 
than compensates for the $1/N_c$ LEC suppression, making the lack of a 
physically constrained estimate for ${\cal C}_1^r$ particularly problematic.

A key point in resolving this problem is the observation that $c_{10}$, 
${\cal C}_0^r$ and ${\cal C}_1^r$ depend differently on $m_{\pi ,K}$. 
Lattice data with variable $m_q$ can thus help disentangle the different 
contributions. Considering the difference of the physical
({\it 'phys'}) and lattice ({\it 'latt'}) versions of 
$\Delta\overline{\Pi}_{V-A}(Q^2)$ {\it at the same $Q^2$}, and implementing 
the NNLO representation for both, yields the new constraints
\begin{eqnarray}
&&\left[\Delta\overline{\Pi}_{V-A}(Q^2)\right]_{latt}-
\left[\Delta\overline{\Pi}_{V-A}(Q^2)\right]_{phys} \, -\, 
\Delta \hat{R}(Q^2)\, =\, \Delta{c_{10}^r}\, L_{10}^r
\, +\, \Delta{c_0}\, {\cal C}_0^r\, +\, \Delta{c_1}\, {\cal C}_1^r\ ,
\label{lattmphysconstraint}\end{eqnarray}
where $\Delta\hat{R}(Q^2)\equiv \left[\hat{R}(Q^2)\right]_{latt}
-\left[\hat{R}(Q^2)\right]_{phys}$, and the expressions
for $\Delta c_{10}$, $\Delta c_0$, $\Delta c_1$ follow from 
those for $c_{10}$, $c_0$ and $c_1$.
%$\Delta c_{10}\, \equiv\, c_{10}^{latt}-c_{10}^{phys}$,
%$\Delta{c_0}=\left( m_{\pi ;latt}^2/m_{\pi ;phys}^2\right)\, -\, 1$ and
%$\Delta{c_1}={\frac{\left(m_\pi^2+2m_K^2\right)_{latt}}
%{\left(m_\pi^2+2m_K^2\right)_{phys}}}\, -\, 1$. 
$\Delta c_{10}$, $\Delta c_0$ and $\Delta c_1$
are all fixed by $\mu$ and the physical and ensemble PS 
mass and $f_\pi$ values; $\Delta\hat{R}(Q^2)$ depends, in addition, 
on $Q^2$. For all ensembles considered here, the combination of 
$L_{10}^r$, ${\cal C}_0^r$ and ${\cal C}_1^r$ in (\ref{lattmphysconstraint}) 
differs significantly from that in (\ref{nnlocontpivma0}). 
Since the RHS of Eq.~(\ref{lattmphysconstraint}) is $Q^2$-independent, 
while all terms on the LHS are $Q^2$-dependent, the constraints
(\ref{lattmphysconstraint}), for a given ensemble, but different $Q^2$, 
provide checks on the self-consistency of the analysis. 

The combination of the lattice constraints (\ref{lattmphysconstraint}) and 
continuum constraint (\ref{nnlocontpivma0}) is sufficient to allow a 
simultaneous determination of $L_{10}^r$, ${\cal C}_0^r$ and ${\cal C}_1^r$, 
albeit with larger-than-ideal errors ($\sim 25\%$, $\sim 100\%$ and 
$\sim 80\%$, respectively), which result from the sizable uncertainties on 
the low-$Q^2$ lattice data. An additional constraint is required to reduce 
these errors.

\begin{figure}[H]
\caption{\label{fig2}$s_0$-dependence of the individual contributions 
and sum on the RHS of the FB V-A IMFESR.}
\begin{center}
\rotatebox{270}{\mbox{
%\begin{minipage}{9.1cm}
\includegraphics[width=7.1cm,height=9.4cm]{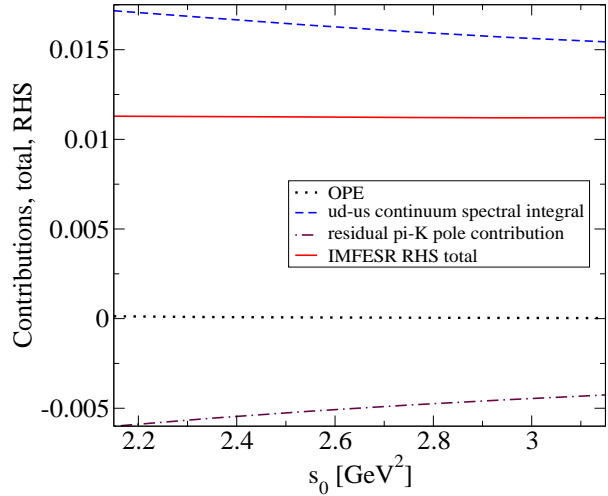}
%\end{minipage}
}}
\end{center}
\end{figure}

Such a constraint can be obtained from inverse moment finite energy sum rules 
(IMFESRs) involving the FB $ud-us$ V-A correlator difference, 
$\Delta\Pi^{FB}_{V-A}\equiv
\Pi^{(0+1)}_{ud;V-A}-\Pi^{(0+1)}_{us;V-A}$~\cite{kmimsr13}. 
Generically, for polynomial $w(s)$, these have the form
\begin{eqnarray}
&&w(0)\, \Delta\Pi^{FB}_{V-A}(0) \, =\, 
{\frac{1}{2\pi i}}\,\oint_{\vert s\vert = s_0} ds\,
{\frac{w(s)}{s}}\, \Delta\Pi^{FB}_{V-A} (Q^2)\, +\, 
\int_0^{s_0}ds\, {\frac{w(s)}{s}}\,\Delta\rho^{FB}_{V-A} (s)\ .
\label{imsr}\end{eqnarray}
The first term on the RHS is to be evaluated using the OPE, the second using 
experimental spectral data. The result is a constraint on the LECs appearing 
in the low-energy representation of the LHS. The choice
$w(s)=w_{DK}(y)\, =\, (1-y)^3\left( 1+y+{\frac{1}{2}}y^2\right)$
with $y=s/s_0$, reduces OPE errors and strongly suppresses spectral 
contributions from the high-$s$ region where uncertainties in the $us$ V/A 
separation are large~\cite{dk00}. With $\Delta\overline{\Pi}^{FB}_{V-A}$
($\Delta\overline{\rho}^{FB}_{V-A}$) the $\pi$- and $K$-pole
subtracted version of $\Delta\Pi^{FB}_{V-A}$ ($\Delta\rho^{FB}_{V-A}$),
$y_\pi =m_\pi^2/s_0$, $y_K=m_K^2/s_0$, and $f_{res}(y)=4y-y^2-y^3-y^4+y^5$,
(\ref{imsr}) can be recast as
\begin{eqnarray}
\Delta\overline{\Pi}^{FB}_{V-A}(0) &&=
{\frac{1}{2\pi i}}\,\oint_{\vert s\vert = s_0} ds\,
{\frac{w_{DK}(y)}{s}}\, \left[\Delta\Pi^{FB}_{V-A} (Q^2)\right]^{OPE} 
\int_{4m_\pi^2}^{s_0}ds\, {\frac{w_{DK}(y)}{s}}\,
\Delta\overline{\rho}^{FB}_{V-A}(s)
\nonumber\\
&&\ \ \ \ \ +\, {\frac{f_\pi^2}{m_\pi^2}}\, f_{res}(y_\pi )\,
-\, {\frac{f_K^2}{m_K^2}}\, f_{res}(y_K)\ .
\label{imsralt}\end{eqnarray}
The $s_0$-independence of the LHS provides a self-consistency test for the 
treatments of the individual $s_0$-dependent terms appearing on the RHS. 
This test is well satisfied, as shown in Fig.~\ref{fig2}~\cite{kmimsr13}.
Strong cancellations, already present in the separate $ud$ and $us$ $D=2,4$ 
V-A series, make the OPE contributions very small. Updated~\cite{dv7} OPAL 
data~\cite{opal99} were employed for the $ud$ spectral integrals. 
For the $us$ spectral integrals, recent B-factory results were used
for $K\pi$~\cite{babarbellekpi}, $K^-\pi^+\pi^-$~\cite{babarkpipiallchg},
and $K_S \pi^-\pi^0$\cite{bellekspipi}, and ALEPH data~\cite{alephus99}, 
rescaled to current branching fractions, for all other modes. The $us$ V/A
separation is unambiguous for $K$ (A) and $K\pi$ (V) contributions, as well 
as for $K^-\pi^+\pi^-$ and $K_S\pi^-\pi^0$ contributions in the $K_1(1270)$ (A)
region. The $1/s$ and $w_{DK}(y)$ weightings (the latter with its 
triple zero at $s=s_0$) mean these contributions dominate the $us$ spectral 
integrals. Higher-$s$ $K\pi\pi$ contributions, and those from all higher 
multiplicity modes, are assigned $50\pm 50\%$ each to the V and A channels, 
with full anticorrelation. 

\unitlength1cm

The use of the combination 
$\Delta\Pi^{FB}_{V-A}$ is predicated on the fact that the NNLO representation
\begin{equation}
\left[ \Delta\overline{\Pi}^{FB}_{V-A}(0)\right]_{NNLO}\, =\, 
R^{FB}_{V-A}(0)\, +\, c^\prime_5 L_5^r\, +\, c^\prime_9 L_9^r
\, +\, c^\prime_{10} L_{10}^r \, -\, 32\left({\frac{m_K^2-m_\pi^2}{m_\pi^2}}
\right)\, {\cal C}_0^r
\label{udmusvmannlorep}\end{equation}
involves a combination of $L_{10}^r$ and ${\cal C}_0^r$ independent of 
those appearing in (\ref{nnlocontpivma0}) and (\ref{lattmphysconstraint}). 
$R^{FB}_{V-A}(0)$, $c^\prime_5$, $c^\prime_9$ and $c^\prime_{10}$ are, as 
usual, determined by $\mu$, $f_\pi$ and the PS masses. Evaluating
all LECs at the conventional chiral scale choice, $\mu =\hat{\mu}
=0.77 \ GeV$, the results of Ref.~\cite{abt00} imply 
\begin{equation}
\left[ \Delta\overline{\Pi}^{FB}_{V-A}(0)\right]_{NNLO}\, =\, 0.00670\,
-\, 0.722 L_5^r \, +\, 1.423 L_9^r\, +\, 2.125 L_{10}^r\,
-\, 11.606 {\cal C}_0^r\ ,
\label{udmusvmannlonumerical}\end{equation}
which, with $L_5^r(\hat{\mu})=0.00058(13)$~\cite{bj11} and
$L_9^r(\hat{\mu})=0.00593(43)$~\cite{bt02},
yields the new constraint
\begin{equation}
2.125 L_{10}^r\, -\, 11.606 {\cal C}_0^r\ ,=
\Delta\overline{\Pi}^{FB}_{V-A}(0)
\, -\, 0.01472\, (61)_{L_9^r}\, (9)_{L_5^r}\ .
\label{vmaimsrfinalconstaint}\end{equation}

\section{\label{sec3}Results}
%Results for $L_{10}^r$, ${\cal C}_0^r$ and ${\cal C}_1^r$ are quoted
%in two stages, first based on the $\Delta\Pi_{V-A}(0)$ and combined 
%lattice/continuum constraints, (\ref{nnlocontpivma0}) and
%(\ref{lattmphysconstraint}), and second for the extended fit incorporating 
%also the IMFESR constraint, (\ref{vmaimsrfinalconstaint}).
%
The results of the continuum $\Delta\overline{\Pi}_{V-A}(Q^2)$ study
of Ref.~\cite{bgjmp13} make clear that the combined lattice/continuum 
constraints can be safely employed only for $Q^2$ up to $\sim 0.3\ GeV^2$. 
For the fine ensembles, this leaves only a few $Q^2$ points with errors 
small enough to allow for meaningful self-consistency tests. One of the 
constraints for the $m_\pi =399\ MeV$ fine ensemble appears clearly 
out of line, but given the limited number of points, we exclude the 
constraints from this ensemble, and base our fits on results from 
the other four ensembles, which display mutually consistent
constraints for all relevant $Q^2$. The results are, in fact,
essentially unchanged if the lack of self-consistency for the
fifth ensemble is ignored, and all ensembles included in the fit.

Results of the $1^{st}$-stage fit, to the $\Delta\Pi_{V-A}(0)$ and combined
lattice-continuum constraints, are
\begin{eqnarray}
&&L_{10}^r(\hat{\mu})\, =\, -0.0031(8),\qquad 
{\cal C}_0^r(\hat{\mu})\, =\, -0.00081(82),\qquad 
{\cal C}_1^r(\hat{\mu})\, =\, 0.014(11)\ .
\label{stage1results}\end{eqnarray}
The non-trivial uncertainties result from the need to determine two 
additional LEC combinations from the relatively large-error lattice data.
Strong correlations among the fit parameters 
mean the additional IMFESR constraint has the possibility 
of improving all three errors.

The input outlined above yields the following result for the RHS of 
Eq.~(\ref{imsralt}):
\begin{equation}
\Delta\overline{\Pi}^{FB}_{V-A}(0)\, =\, 0.01125\, (135)_{OPAL}\, (16)_{res}
\, (15)_{OPE}\, (5)_{s_0}\ .
\label{udmusvmapi0value}\end{equation}
The subscripts {\it 'OPAL', 'res', 'OPE' and '$s_0$'} label errors
associated with uncertainties in the OPAL continuum
distributions, residual $\pi$- and $K$-pole and OPE contributions, and 
the (very small) residual $s_0$-dependence, respectively.
Adding the resulting IMFESR constraint to the combined fit yields
\begin{eqnarray}
&&L_{10}^r(\hat{\mu})\, =\, -0.00346(29),\qquad
{\cal C}_0^r(\hat{\mu})\, =\, -0.00034(12),\qquad 
{\cal C}_1^r(\hat{\mu})\, =\, 0.0081(31)\ .
\label{finalfitresults}\end{eqnarray}
The result for $L_{10}^r$ is the most precise to date, with, moreover,
the errors purely data-driven. The result for ${\cal C}_1^r(\hat{\mu})$ lies 
significantly outside the range assumed in Ref.~\cite{gapp08}. Its non-zero
value also provides a further example of an LEC combination which vanishes in 
the large-$N_c$ limit, but cannot be neglected for $N_c=3$. The difference 
between our result for ${\cal C}_0^r(\hat{\mu})$ and that employed in 
Ref.~\cite{gapp08}, $0.00054(42)$, has two sources. The first is 
a shift in $C_{61}^r$ due to significant shifts in the input to the analysis 
underlying the original $C_{61}^r$ determination~\cite{dk00}, the second 
the RChPT result for $C_{80}^r$ used in Ref.~\cite{gapp08}, which turns out
to represent a significant overestimate of the true 
value~\cite{kmimsr13}.

\section{Acknowledgements}
The lattice computations were done using the STFC's DiRAC facilities at Swansea
and Edinburgh. PAB, LDD, NG and RJH are supported by an STFC Consolidated
Grant, and by the EU under Grant Agreement PITN-GA-2009-238353 (ITN
STRONGnet). EK was supported by the Comunidad Aut\`onoma de Madrid under
the program HEPHACOS S2009/ESP-1473 and the European Union under Grant
Agreement PITN-GA-2009-238353 (ITN STRONGnet).
KM acknowledges the hospitality of the CSSM, University of Adelaide and
IFAE Barcelona, and the support of NSERC (Canada).
JMZ is supported by the Australian Research Council grant FT100100005.

\end{document}